\documentclass[
    ,final            % use final for the camera ready runs
  ]
  {aipproc}

\layoutstyle{8x11double}

\usepackage{amssymb}

\begin{document}

\title{Neutron Stars: Formed, Spun and Kicked}


\author{V. Kalogera}{
  address={Northwestern University, Department of Physics and Astronomy,  2131 Tech Drive, Evanston, IL 60208, USA}
}

\author{F. Valsecchi}{
  address={Northwestern University, Department of Physics and Astronomy,
  2131 Tech Drive, Evanston, IL 60208, USA}
}

\author{B. Willems}{
  address={Northwestern University, Department of Physics and Astronomy,
  2131 Tech Drive, Evanston, IL 60208, USA}
}

%\author{K. Belczynski}{
%  address={New Mexico State University, Department of Astronomy, 1320 Frenger Mall, Las Cruces, NM 88003, USA}
%  ,altaddress={Tombaugh Fellow} % additional visiting address
%}

\begin{abstract}

One of the primary goals when studying stellar systems with neutron stars has been to reveal the physical properties of progenitors and understand how neutron star spins and birth kicks are determined. Over the years a consensus understanding had been developed, 
% favoring Fe core collapse of He-rich stars more massive than 2\,M$_\odot$, birth kicks of 200--300\,km\,s$^{-1}$ and spin up through stable mass accretion driven by Roche-lobe overflow. 
but recently some of the basic elements of this understanding are being challenged by current observations of
some binary systems and their theoretical interpretation. In what follows we review such recent developments and highlight how they are interconnected; we particularly emphasize some of the assumptions
and caveats of theoretical interpretations and examine their validity (e.g., in connection to the unknown radial velocities of pulsars or the nuances of multi-dimensional statistical analysis). The emerging picture does not erase our earlier understanding; instead it broadens it as it reveals additional pathways for neutron star formation and evolution: neutron stars probably form at the end of both core collapse of Fe cores of massive stars and electron-capture supernovae of ONeMg cores of lower-mass stars; birth kicks are required to be high (well in excess of 100\,km\,s$^{-1}$) for some neutron stars and low ($<100$\,km\,s$^{-1}$) for others depending on the formation process; and spin up may occur not just through Roche-lobe overflow but also through wind accretion or phases of hypercritical accretion during common envelope evolution. 

\end{abstract}

%%%%%%%%%%%%%%%%%%%%%%%%%%%%%%%%%%%%%%%%%%%%%%%%%%%%%%%%%%%%%%%%%%%
%%
%% The below \maketitle command inserts the actual front matter data.
%% It has to follow the above declarations.
%%
%%%%%%%%%%%%%%%%%%%%%%%%%%%

\maketitle

%%%%%%%%%%%%%%%%%%%%%%%%%%%%%%%%%%%%%%%%%%%%
%% MAINMATTER
%%
%%%%%%%%%%%%%%%%%%%%%%%%%%%%%%%%%%%%%%%%%%%%%%%%%%%%%%%%%%%%%%%%%%%%%%%%%%%%
%% Headings:
%%
%% The aipproc supports three heading levels, i.e., \section,
%%	\subsection, and \subsubsection.
%%%%%%%%%%%%%%%%%%%%%%%%%%%%%%%%%%%%%%%%%%%%%%%%%%%%%%%%%%%%%%%%%%%%%%%%%%%%
%% Cross-references:
%%
%% Page numbers (\pageref) and headings can NOT be referenced in the class,
%% since before being produced, no page numbers are determined.
%%
%% Tables, figures, and equeations can be referenced by using the LaTex
%% 	commands \label and \ref. For references to equation numbers, \eqref
%%	can be used, which will print "(1)" (while \ref will result in "1").
%%
%%%%%%%%%%%%%%%%%%%%%%%%%%%%%%%%%%%%%%%%%%%%%%%%%%%%%%%%%%%%%%%%%%%%%%%%%%%%
%% Lists: 
%%
%% Standard "itemize", "enumerate", etc. list environments are supported.
%%%%%%%%%%%%%%%%%%%%%%%%%%%%%%%%%%%%%%%%%%%%%%%%%%%%%%%%%%%%%%%%%%%%%%%%%%%%
%% Urls:
%%
%% \url{} command is provided for documenting URLs.
%%%%%%%%%%%%%%%%%%%%%%%%%%%%%%%%%%%%%%%%%%%%

\section{Neutron Star Formation}

\subsection{Elements of a ``Standard Picture''} 

Forty years after the discovery of pulsars it is fair to say that an immense amount of progress has been made in the quest to understand how neutron stars (NS) form and evolve in different stellar systems. Our understanding of these processes has benefited most importantly by the study of binary pulsars or X-ray binaries harboring NS with either stellar or degenerate companions (white dwarfs or other NS). Progress has been made both by studying specific observed systems as well as whole populations in the Galactic field or in globular clusters.These analyses and results have always coupled strongly to efforts in the areas of core-collapse and supernova (SN) explosions of massive stars, thermonuclear explosions of accreting white dwarfs, and nuclear evolution of massive stars. 

Through this diverse body of work (for which it is impossible to include here all important references, but relevant overviews may be found in \citep{2006csxs.book..623T, 1991PhR...203....1B}) a generally accepted understanding seemed to have emerged anchored on the following crucial elements most relevant to the discussion presented in this paper: (i) NS form at the end of stellar nuclear evolution as remnants of SN explosions of massive stars of $\sim 10$\,M$_\odot$ up to $\sim 20-40$\,M$_\odot$; (ii) NS in close binaries typically have helium-rich progenitors at the time of core collapse; these progenitors are the helium-rich cores (typically more massive than 2\,M$_\odot$) of original massive binary companions which have lost their hydrogen-rich envelopes through binary interactions involving (stable or unstable) mass transfer between binary components and/or mass loss from the binary\footnote{In binaries with low-mass companions NS may also form through the accretion-induced collapse of white dwarfs, but here we focus on massive binaries.}; (iii) at the time of their formation NS acquire recoil kicks due to asymmetries that develop during the core-collapse process; typical kick magnitudes inferred from studies of pulsar population kinematics and the formation of binaries with NS are in the range of 100--300\,km\,s$^{-1}$; (iv) NS are born with birth spin periods of $\sim10$\,ms and spin down over a time span of $\sim 10$\,Myr; if in binaries they can spin up and reach spin periods of 10\,ms or shorter later in their life through accretion of material from their companion and accompanied reduction in magnetic-field strength.  

\subsection{Recent Developments and Challenges} 

In recent years a growing number of NS studies have challenged different aspects of the ``standard picture'' as described above. Although each one of these in isolation may not have been powerful enough to cause a paradigm shift, the current set of inter-linked arguments presented in the literature in the past $\sim 5$ years has certainly attracted attention and  has opened the door to potential changes in our current understanding of how NS form. In fact, as we will discuss in what follows, these recent developments do not actually erase or substitute our earlier understanding; instead they bring forward new possibilities that had not received much attention or had not even been raised previously.  

With a broad-brush look, two main physical processes lie at the core of a lot of the issues raised in connection to NS formation and evolution: (i) {\em electron-capture supernovae} (ECS), and (ii) {\em hypercritical accretion onto NS} in common envelope phases (dynamically unstable mass transfer leading to common-envelope formation and ejection through NS inspiral towards the core of a massive companion). 

The possibility of ECS was originally put forward as a potential channel for NS formation that does not involve the ``usual'' collapse of the iron core of a massive star at the end of nuclear evolution \citep{1980PASJ...32..303M,1984ApJ...277..791N,1987ApJ...322..206N}. Instead massive ONeMg white dwarfs can reach high enough densities for electrons to capture onto $^{24}$Mg and $^{20}$Ne (before Ne ignition can occur (see \citep{1980PASJ...32..303M, 1998ApJ...493L.101W, 2003ApJ...593..968W}),  causing the white dwarf to lose support against gravity; ultimately this process is expected to lead to core collapse, a SN, and a NS remnant. Such a collapse could occur right at the progenitor mass boundary between the formation of massive, ONeMg white dwarfs and the collapse of iron cores into NS: hydrogen-rich, original progenitors of $\sim 8-10$\,M$_\odot$. This mass range is, however, highly uncertain. 

More recently Podsiadlowski et al.\ \citep{2004ApJ...612.1044P} proposed that ECS events are relevant for stars of initial mass in the range $8-11$\,M$_\odot$ in binary systems where they can avoid the second dredge-up phase on the asymptotic giant branch (AGB) (due to loss of the hydrogen-rich envelope) and form bigger helium cores; instead single stars of the same mass would form ONeMg white dwarfs. They also suggested that ECS are relatively fast explosions compared to delayed neutrino-driven explosions of iron cores and could naturally lead to natal kicks of low magnitudes. Kitaura et al.\ \citep{2006A&A...450..345K} modeled ECS explosions of a $\simeq$1.3\,M$_\odot$ ONeMg white dwarf (formed from a 2.2\,M$_\odot$ helium core, which in turn formed from an 8.8\,M$_\odot$ hydrogen-rich progenitor); they find that the explosions are rather faint (low $^{56}$Ni production), produce low-mass NS with baryonic mass of $1.36$\,M$_\odot$, and are not prompt but still delayed and powered by neutrino heating.  T.\ Janka during his talk at this conference \citep{J2008} presented more recent results and commented that the shock is not delayed long enough for kicks in excess of $\sim 100$\,km\,s$^{-1}$ to be produced. Last, Poelarends et al.\ \citep{2007arXiv0705.4643P} investigated the late evolution of solar-metallicity {\em single} stars on the AGB and concluded that stars in the mass range $9-9.25$\,M$_\odot$ could experience ECS, implying an ECS fraction of 4--20\% among all core-collapse SN, if one accounts for some of the model uncertainties (AGB stellar wind strengths and dredge-up efficiency, semi-convection, convective overshooting, rotational mixing; see \citep{2004MNRAS.353...87E, 2007A&A...476..893S}). In a similar study, Siess \citep{2007A&A...476..893S} concluded that the ECS mass range is indeed very narrow for single stars (<1--1.5\,M$_\odot$); at solar metallicity he finds a mass range of 9.5--10.5\,M$_\odot$ which is shifted downwards by 2\,M$_\odot$ if core overshooting is included and could also vanish altogether depending on the metallicity dependence of AGB winds.    

In view of all this recent work, our current understanding of ECS in connection to NS progenitors and kicks can be summarized as follows: they can occur in nature and may contribute a small but non-negligible fraction of NS forming with natal kicks $\lesssim 100$\,km\,s$^{-1}$. Their progenitor mass range in single and binary stars is {\em highly uncertain}, but {\em may be as wide as} $7-12$\,M$_\odot$ \citep{2004ApJ...612.1044P, 2007arXiv0705.4643P, 2007A&A...476..893S}, and even reach down to 5\,M$_\odot$ for low-metallicity stars \citep{2007A&A...464..667G}, potentially including helium stars less massive than 2\,M$_\odot$ \citep{1980PASJ...32..303M, 1987ApJ...318..307M, 1993ApJ...414L.105H}. 

A significant part of our understanding of how NS form and evolve comes from analyses of binary pulsars with two neutron stars (NS-NS). 
In the ``standard'' NS-NS formation channel \citep{1991PhR...203....1B} the first-born NS experiences a common-envelope phase when its hydrogen-rich companion fills its Roche lobe. As a result the orbit shrinks to the small orbits observed, the helium-rich progenitor of the second NS is revealed, and the first NS has a chance of getting mildly recycled due to accretion in the common envelope and/or mass transfer from the helium star companion. Brown \citep{1995ApJ...440..270B} challenged the survival of the first NS through the common-envelope phase and argued that the NS will collapse instead into a black hole due to hyper-critical accretion (exceeding the photon Eddington limit by many orders of magnitude). He suggested an alternative formation scenario that envisions evolution through a double-common-envelope phase revealing two helium-rich NS progenitors  in a tight orbit, followed by two consequent SN explosions. This channel requires that the two hydrogen-rich massive progenitors have almost equal masses (within 4\%) and could also potentially explain the almost equal NS masses in NS-NS. We note however, that hydrodynamical studies of hypercritical accretion (e.g., \citep{1996ApJ...460..801F}) and population studies of NS-NS formation that have accounted for hypercritical accretion (e.g., \citep{2002ApJ...572..407B}) have concluded that if the maximum NS mass is $\simeq 2$\,M$_\odot$ or higher, then a significant fraction of NS can still survive the common-envelope phase and avoid collapse into a black hole.

\section{Progenitor Masses}

For almost two decades now, it has been widely accepted that there is a minimum mass a helium-rich star must have for the formation of an iron core and NS formation at core collapse. This minimum value is somewhat dependent on the detailed treatment of convection and late phases of nuclear burning in stellar models, but has been known to be in excess of 2\,M$_\odot$ (e.g., in the range 2.1--2.3\,M$_\odot$, see Habets \citep{1986A&A...167...61H}).  

One pathway for helium stars to be less massive than the above limit at the time of core collapse is provided by mass transfer onto its binary companion, which however does not abort iron-core formation. Recently, several independent groups have analysed the late stages of evolution of NS-NS progenitors and all reached the same conclusion:  at least two of the known close NS-NS binaries (PSR J0737-3039 and PSR~B1534+12) were experiencing Roche-lobe overflow mass transfer from the helium-rich progenitor of the second NS onto the first NS at the time of the second SN explosion \citep{2004MNRAS.349..169D, 2004ApJ...603L.101W, 2004ApJ...616..414W, 2005ApJ...619.1036T, 2006MNRAS.373L..50S}.
Although the mass accretion onto the NS may be rather small (and yet still adequate for mild recycling), the mass of the helium star can be reduced significantly ($\simeq 0.5 - 1.5$\,M$_\odot$ from its original mass by the time of its core collapse \citep{2002MNRAS.331.1027D, 2003ApJ...592..475I, 2003MNRAS.344..629D}. 

Alternatively (or even concurrently?) NS formed in ECS are also expected to originate from low-mass helium cores. This is primarily because of the connection to an ONeMg core that just barely exceeds the Chandrasekhar mass when electron capture and loss of support ensue. Although the ECS progenitor mass range is highly uncertain (see previous section), it tends to lie below 11--12\,M$_\odot$ and potentially includes helium-rich progenitors less massive than 2\,M$_\odot$ (see ECS discussion in \citep{2005astro.ph.11811B, 2007arXiv0706.4096I}. Such low-mass progenitors would also naturally produce low-mass NS with baryonic masses very close to the Chandrasekhar limit and gravitational masses of $\simeq 1.25$\,M$_\odot$ \citep{2006A&A...450..345K, 2007arXiv0706.4096I}.  

The discovery of the double pulsar \citep{2003Natur.426..531B} and the mass measurement of pulsar B (the second-born NS) at 1.25\,M$_\odot$  led Podsiadlowski et al.\ \citep{2005MNRAS.361.1243P} to suggest an ECS origin. A number of independent groups analyzed the evolutionary history of this system, given the observed characteristics, and derived constraints on the immediate (right before the SN) progenitor of pulsar B. Piran \& Shaviv \citep{2005PhRvL..94e1102P} were the first to allow as part of the priors in the statistical analysis helium star progenitors less massive than 2\,M$_\odot$; they found that progenitor masses in the range 1.25--1.75\,M$_\odot$ were favored by their analysis at the 30\% confidence level, while masses up to 2\,M$_\odot$ could just be reached 
at 97\% confidence level (see their Figure 2). Willems et al.\ \citep{2006PhRvD..74d3003W} presented a detailed analysis that explored the influence of all uncertainties associated with the systems (most notably its age and radial velocity with respect to the Sun); when allowing for low-mass progenitors they too concluded that the {most likely} helium star masses lie in the range 1.35--1.95\,M$_\odot$, while the 90\% confidence intervals extended up to 2.5\,M$_\odot$ (see their Figure 6 and Table IV). More recently, Stairs et al.\ \citep{2006MNRAS.373L..50S} addressed some of these questions using updated proper motion measurements, and they too concluded that the immediate progenitor of pulsar B was likely less massive than 2\,M$_\odot$ with most likely values at $\simeq 1.5$\,M$_\odot$ (see their Figure 1). 
These conclusions are very different from those of a very similar analysis of the NS-NS system PSR~B1534+12 for which 
%even when no prior lower limit on the helium progenitor mass of the second-born NS is set: 
Stairs et al.\ \citep{2006MNRAS.373L..50S} report a most likely value of 2.45\,M$_\odot$ and a 95\% confidence interval 1.6--3.9\,M$_\odot$. 

Based on these results, it appears most probable that the immediate progenitor of pulsar B was less massive than 2\,M$_\odot$ (most likely $\simeq 1.5$\,M$_\odot$), even though this is clearly not universal among close NS-NS binaries. The next question becomes: {\em what does this low mass tell us about NS formation? } We would argue that the answer is still unclear. It is indeed very tempting to associate this mass with an ECS event and declare the case closed, and this could very well be the answer. However, we should not eliminate from our thinking that PSR~J0737-3039 has also been found by several independent studies to be in a phase of Roche-lobe  at the time of pulsar B formation. Consequently, we cannot exclude that the low progenitor mass is due to mass transfer (albeit with small accretion onto pulsar A) initiated before the SN explosion. {\em We conclude that both of these possibilities remain as potentially relevant to the double pulsar formation.}  

At present no obvious secure observational test to distinguish between these two possibilities appears to exist. However we suggest that two different types of theoretical studies could shed some light and allow us to make further progress: {\em (i)} detailed mass transfer sequences that can explain not just the low progenitor mass but also the orbital characteristics right at the explosion time as well as the measured NS masses; {\em (ii)} population studies simulating evolution forward in time focusing on the question of the relative frequency of ECS vs. mass transfer effects depending on assumptions about the highly uncertain ECS progenitor mass range.

%Nomoto 1984, 1987, Pols et al. 1998, Hurley et al. 2000, Podsiadlowski et al. 2004, Eldridge et al. 2006, Ivanova et al. 2007, Poelarends et al. 2007

\section{Supernova Kick Magnitudes}

For over more than three decades now, estimates of kick velocity magnitudes imparted to NS at birth have been obtained through the analysis of the kinematics of single pulsar samples of varying sizes and reliability of proper motion measurements (starting with Gunn \& Ostriker \citep{1970ApJ...160..979G}).  Arzoumanian et al.\ \citep{2002ApJ...568..289A} analyzed a sample of 79 single pulsar proper motions and concluded that the NS kick magnitude distribution is best described by a two-component Maxwellian velocity distribution with characteristic velocities (three-dimensional velocity dispersions) of 90\,km\,s$^{-1}$ and 500\,km\,s$^{-1}$ with almost equal relative weight. More recently, Hobbs et al.\ \citep{2005MNRAS.360..974H} adopted an updated electron density model and used a sample of 73 proper motion measurements for 
pulsars with short characteristic ages ($\lesssim 3$\,Myr); they concluded that the sample is best described by a single Maxwellian with a one-dimensional velocity dispersion of 265\,km\,s$^{-1}$, shifting the mean kick velocity to values $\simeq 400$\,km\,s$^{-1}$. Very similar results were also obtained by Zou et al.\ \citep{2005MNRAS.362.1189Z}. It is recognized that these results could be somewhat contaminated by pulsars originating from disrupted binaries and the contribution of orbital velocities, and by a potential bias against low velocities due to the associated small and hence not easily measurable proper motions. Nevertheless, it is clear that overall these analyses point to significant kick magnitudes of a few hundreds of km\,s$^{-1}$. 

Additional constraints on NS kick magnitudes have been obtained by the analysis of the evolutionary history of specific binary systems with NS known in the Galaxy, most notably binary pulsars and X-ray binaries. Although the majority of such studies in the past reached conclusions in support of significant NS kicks, more recently a number of arguments and analyses point towards a need for at least some NS receiving low-magnitude kicks. In what follows we summarize such evidence and place it in the context of our developing current understanding. 

{\bf Neutron Stars in Globular Clusters.} Perhaps one of the earliest arguments in favor of modest or negligible NS birth kicks has been anchored on the need for NS retention in globular clusters. The observed low-mass X-ray binaries (persistent and transient) and binary pulsars in Galactic globular clusters have led to estimates for the total number of NS currently present in clusters in the range $300-1,000$ \citep{2005ApJ...625..796H}, implying retention fractions of $1-10$\%. Given that current estimates of cluster escape velocities lie in the range $\simeq 20-50$\,km\,s$^{-1}$, it is evident that clusters could indeed pose a requirement for low kick magnitudes. Pfahl et al.\ \citep{2002ApJ...573..283P} concluded that retention fractions of 1--10\% can be achieved if kick magnitude distributions are Maxwellians with one-dimensional dispersions smaller than $\simeq 200$\,km\,s$^{-1}$. They also found that the vast majority of NS are retained in massive binaries in the clusters. However the increased characteristic kick magnitudes derived by Hobbs et al.\ \citep{2005MNRAS.360..974H} re-emphasized the issue of NS retention in clusters. Most recently, Ivanova et al.\ \citep{2007arXiv0706.4096I, 2006MNRAS.372.1043I} and Kuranov \& Postnov \citep{2006AstL...32..393K} 
concluded that the required retention fractions can still be obtained, if a significant fraction of NS form through ECS events, and if these NS acquire either no kicks or kicks reduced by a factor of 10 relative to kicks imparted to NS formed through iron core collapse. 

{\bf High Mass X-ray Binaries.} Pfahl et al.\ \citep{2002ApJ...574..364P} highlighted the existence of a  subclass of high-mass X-ray binaries characterized by relatively low orbital eccentricities ($<0.2$) and wide orbits (periods in excess of 30\,d). They argued that their formation appears highly unlikely, unless some NS acquire only small kicks of $\simeq20$\,km\,s$^{-1}$ (see their Figures 3 and 5). Subsequently Podsiadlowski et al.\ \citep{2004ApJ...612.1044P} associated these low kicks with NS formation in ECS events and accounted for the subclass of low-eccentricity, wide high-mass X-ray binaries assuming a rather wide ECS mass range of $8-14$\,M$_\odot$. 

{\bf Double Neutron Stars.} The pattern of observed systems requiring both low and high NS kick magnitudes appears to be even more perplexed for the case of NS-NS binaries. 
To our knowledge van den Heuvel \citep{2004ESASP.552..185V} was the first to point to NS-NS and argue in favor of low NS kicks based on the occurrence of small eccentricities ($<0.3$) for 5 of the 7 systems known in the Galactic field. Following Pfahl et al.\ \citep{2002ApJ...573..283P} and Podsiadlowski et al.\ \citep{2004ApJ...612.1044P}, he suggested that {\em some} second-born NS must have received very small kicks ($\simeq 20$\,km\,s$^{-1}$) because of ECS formation. On the other hand, Chaurasia \& Bailes \citep{2005ApJ...632.1054C} and Ihm et al.\ \citep{2006ApJ...652..540I} showed that orbital evolution through gravitational radiation alters the eccentricity distribution: many of the high-eccentricity systems are drained from the population because they evolve fast and either merge and/or shift to lower eccentricities through circularization. 

McLaughlin et al.\ \citep{2005ASPC..328...43M} highlighted a correlation between the NS-NS eccentricities and the spin periods of the first-born NS in the systems, while Faulkner et al.\  \citep{2005ApJ...618L.119F} argued that such a correlation would naturally arise if all second-born NS were formed symmetrically and no kicks were imparted to them. Following this initial suggestion, Dewi et al.\ \citep{2005MNRAS.363L..71D} performed population studies adopting a simple spin-up model for the first-born NS and concluded that such a correlation appears only if {\em all} second-born NS acquire kicks from a Maxwellian with a velocity dispersion of $\lesssim 50$\,km\,s$^{-1}$. Willems et al.\ \cite{W2008} explored the potential role of polar kicks in forming this correlation and similarly conclude that low kicks for the second-born NS seem to be needed in order to explain this correlation.  
 
Not surprisingly, the particular case of the {\em double pulsar} (PSR~J0737-3039) has attracted a lot of attention in this discussion of NS kick magnitudes. The binary pulsar has a low eccentricity (0.09) and a low second-born NS mass (1.25\,M$_\odot$), and most recently a low proper motion (10\,km\,s$^{-1}$) was reported \citep{2006Sci...314...97K}. Hence, it appears as the prime candidate for a system where the second-born NS was formed through an ECS event and acquired a small kick. 
Willems et al.\ \citep{2004ApJ...616..414W} assumed the ``standard'' minimum progenitor mass of 2\,M$_\odot$ and adopted a flat prior distribution for the unknown radial velocity and found most probably kick magnitudes in the range 100--300\,km\,s$^{-1}$. 
Podsiadlowski et al.\ \citep{2005MNRAS.361.1243P} used the low mass of pulsar B as the ``tell-tale sign'' for ECS and hence a small kick imparted to it, anticipating a small proper motion measurement. Piran \& Shaviv \citep{2005PhRvL..94e1102P} 
used the measured orbital characteristics and an earlier estimate of the proper motion to derive a probability distribution for the birth kick to pulsar B. In their calculation they allowed for the possibility of very low-mass progenitors and also made a number of restrictive assumptions about the system's age and its Galactic motion (the latter allowed them to by-pass the question of the unknown radial velocity). They primarily highlighted their result derived at 30\% confidence level favoring kick magnitudes either $<30$\,km\,s$^{-1}$ or in the range $50-80$\,km\,s$^{-1}$. At a safer confidence level of 97\%, their results set an {\em upper limit of} 125\,km\,s$^{-1}$ (see their Figure 2C).  

Following these studies, Willems et al.\ \citep{2006PhRvD..74d3003W} and Stairs et al.\ \citep{2006MNRAS.373L..50S} presented the two most comprehensive analyses of this system to date (the latter of the two using the most recent and reliable measurement from pulsar timing of the proper motion by Kramer et al.\ \citep{2006Sci...314...97K}). Both reached rather similar conclusions for the {\em most likely} kick magnitude imparted to pulsar B: 50--170\,km\,s$^{-1}$ (Willems et al. \citep{2006PhRvD..74d3003W}) and 40--80\,km\,s$^{-1}$ (Stairs et al. \citep{2006MNRAS.373L..50S}). As expected these magnitudes reach much higher values if wider confidence intervals are considered. Willems et al.\ \citep{2006PhRvD..74d3003W} found up to 300\,km\,s$^{-1}$ at 90\% confidence level, assuming a Gaussian distribution for the unknown radial velocity with dispersion of 130\,km\,s$^{-1}$ (motivated by population synthesis calculations). Stairs et al.\ \citep{2006MNRAS.373L..50S} found kicks in the range of 45--1000\,km\,s$^{-1}$ for the same radial velocity  assumptions, and of 20--140\,km\,s$^{-1}$, assuming velocity vectors are isotropically  distributed around the transverse velocity. The issue of the radial velocities has been identified by these two studies in particular as quite important for the best possible determination of the kick magnitude. To avoid a long digression we discuss this issue in more detail in the Appendix where we show the simple assumption of isotropicity for the system velocity to not be justified. 

It is interesting to compare the Willems et al.\ \citep{2006PhRvD..74d3003W} and Stairs et al.\ \citep{2006MNRAS.373L..50S} results to those obtained by Piran \& Shaviv \citep{2005PhRvL..94e1102P}. Although each study has used somewhat different observational constraints as they became available, one can say that at zeroth order all three studies agree that, once immediate progenitor masses lower than 2\,M$_\odot$ are allowed as part of the priors in the statistical analysis (motivated by ECS events or collapse of rather small iron cores), then kick magnitudes in the range $10-100$\,km\,s$^{-1}$ are clearly favored at low confidence intervals ($\simeq 30$\%). For these magnitudes, Willems et al.\ \citep{2004ApJ...616..414W} and Stairs et al. \citep{2006MNRAS.373L..50S} predicted a spin misalignment angle for pulsar A smaller than 20$^{\rm o}$ in very good agreement with the upper limit of 10--15$^{\rm o}$ at similar confidence intervals just recently reported by Ferdman et al.\ \citep{F2008}. 

Some subtleties however do arise at a more detailed level of examination. Although these do not alter the basic conclusions in any way, it is somewhat important to briefly identify some of them, to avoid propagation to misunderstandings or misconceptions. For example: (i) a NS progenitor of mass lower than 2\,M$_\odot$ does not necessarily imply that kicks must also be very small ($\simeq 20$\,km\,s$^{-1}$); (ii) the current position of the system close to the Galactic plane does not necessarily imply that the system and kick velocity must be small (see the point made by Figure 9b in \citep{2006PhRvD..74d3003W}); (iii) the derived probability distributions are inherently multi-dimensional (5-D: progenitor mass, orbital separation, kick magnitude, and two kick direction angles) and statements about preferred kick magnitudes are {\em quite sensitive} to projection effects in different dimensions because of the complex structure of the multi-dimensional surface (as shown in Table III in \citep{2006PhRvD..74d3003W}, the statement of ``{\em most likely}'' kick can vary by a factor of $\sim 7$ from 10 to 75\,km\,s$^{-1}$ depending on the projection dimensionality); (iv) the system age is one of the poorest constrained priors in the determination of the probability distributions (as seen from Table III in Willems et al. \citep{2006PhRvD..74d3003W}, the ``{\em most likely}'' kick can vary by a factor of $\sim 30$ from 5 to 170\,km\,s$^{-1}$ depending on the assumed system age); (v) the most likely birth location of the system can be several kpc away from its current position (see Figure 8 in \citep{2006PhRvD..74d3003W}), so that following the three dimensional motion of the binary is important for the dynamics.

The results obtained for the {\em other NS-NS binaries} are particularly interesting in comparison to the case of the double pulsar. Just like with the second-born NS progenitor masses, PSR~B1534+12 paints a rather different picture for NS kicks. Independent studies \citep{2004ApJ...616..414W, 2005ApJ...619.1036T} have concluded that a minimum kick of $\simeq200$\,km\,s$^{-1}$ is needed to explain its properties along with its significant spin-orbit misalignment. It is further interesting to point out that this system has a rather small eccentricity (<0.3) {\em and} exhibits rather well the correlation between pulsar spin period and orbital eccentricity. Both of these attributes have been associated  with second-born NS getting no or minimal ($\simeq 20$\,km\,s$^{-1}$) kicks \citep{2004ESASP.552..185V, 2005MNRAS.363L..71D, W2008}. PSR~1534+12 therefore clearly presents a contradictory example. PSR~B1913+16 also clearly requires a significant kick magnitude of at least 260\,km\,s$^{-1}$ \citep[][and references therein]{2004ApJ...616..414W, S2008}.

\section{Spin Up Opportunities for Double Neutron Stars} 

It has been long recognized that first-born NS in NS-NS must have been spun-up (recycled) at some stage prior to the second NS formation event. However, the measured spin periods are all rather long in the range 10--100\,ms, in excess of the ``classical'' millisecond pulsars, and these NS are thought to be only {\em mildly} recycled. Nevertheless, even mild recycling {\em requires} some level of accretion onto the NS at some stage between the two SN events. This requirement is sometimes ignored in discussions of NS-NS formation, but can provide us with some insights and constraints. 

In the ``standard'' NS-NS formation channel, the first-born NS has in principle two opportunities for recycling: either during the common-envelope inspiral involving the hydrogen-rich progenitor of the second NS or during a mass-transfer phase driven by Roche-lobe overflow from the helium core exposed after the common-envelope phase. As mentioned already, the progenitors of at least two of the known NS-NS binaries were indeed in such a mass transfer phase from the helium-rich companion to the first-born NS at the time of the second SN explosion. Therefore the need for NS spin-up during the common envelope is alleviated for these systems. 

It is also interesting that all explanations of the correlation between spin period and eccentricity rely on a (linear) scaling between the accretion amount responsible for the spin-up and the mass of the second-born NS progenitor right before the explosion. Such a scaling arises naturally if spin-up happens during a phase of Roche-lobe overflow mass transfer from a helium-rich companion and is supported by mass-transfer calculations of NS helium-star binaries \citep{2002MNRAS.331.1027D, 2003ApJ...592..475I, 2003MNRAS.344..629D} (see also Fig. 2 in \citep{W2008}). Willems et al. \citep{W2008} furthermore noted that the presence of high-eccentricity systems on the spin period-eccentricity relation could support the existence of a similar scaling for NS accretion during common envelope phases.

Last, it is important to realize that in the case of the double common-envelope channel put forward by Brown \citep{1995ApJ...440..270B}, the opportunities for NS spin-up are limited and potentially negligible. Given how similar the masses of the two progenitors are, the time between the two SN explosions is very short ($<1,000$\,yr) and it is difficult to imagine that in this short time both the orbital and stellar characteristics would happen to be appropriate for establishing Roche-lobe overflow and spinning up the first-born NS. In their population simulations Belczynski et al.\ \citep{2002ApJ...572..407B} did indeed identify this NS-NS formation channel, but, for a maximum NS mass of 3\,M$_\odot$, it only contributed 8\% of the total NS-NS population. Most importantly though Belczynski et al. did not find any opportunity for NS recycling. They also considered a model with a maximum NS mass of 1.5\,M$_\odot$ (model D2) and naturally found a significant contribution to NS-NS formation from the double common-envelope channel, but the issue of recycling remained. More recently Dewi et al.\ \citep{2006MNRAS.368.1742D} modeled NS-NS formation through this channel and obtained rates very similar to those by Belczynski et al.\ \citep{2002ApJ...572..407B}, but did not comment on the difficulty with recycling the first-born NS in the systems. 

At present we would say that the question of when and under what circumstances first-born NS in NS-NS binaries get recycled is rather unclear. If the double common-envelope channel is shown to be dominant in some way, then spin-up during mass transfer from the helium-rich companion would be necessary. It is possible that accretion during common envelope inspiral or wind accretion alone without Roche-lobe overflow would be sufficient to recycle the first born NS, although so far neither of these have been examined quantitatively. Also, if the correlation between spin period and orbital eccentricity persists, the amount of mass accreted by the NS must scale with the donor mass regardless of the mode of mass transfer.

\section{Current Conclusions: One Size Does Not Fit All} 

The picture of NS formation from 10 years ago has clearly changed and paradigms have shifted - NS formation now appears more diverse than we thought both in terms of progenitor masses and birth kicks, and thus also in terms of core-collapse physics, asymmetry developments, and maybe even kick mechanisms. 

In the last 3 years, the number of papers favoring small or even zero NS kicks has mushroomed. A lot of this recent work is driven by the increasing observational information regarding binary pulsars as well as new detections. We summarize the current picture as follows: 
\noindent
(1) NS form from the iron core collapse of  helium-rich
progenitors with masses above $\simeq 2\,M_\odot$, though lower masses may be possible if significant mass loss through Roche lobe overflow occurs shortly before NS formation. Alternatively, NS form from lower-mass helium stars through ECS of massive ONeMg cores. The ECS progenitor mass range is currently highly uncertain, but the channel may be particularly accessible for NS formation in binaries where the NS progenitors lose their hydrogen-rich envelope before the second dredge-up on the AGB. 
\noindent
(2) Typical NS kicks of 200-300\,km\,s$^{-1}$ are still favored by analyses of pulsar kinematics and are required for many of the known binary NS systems. However sub-populations of or some specific NS binaries clearly favor smaller kick magnitudes (lower than $\sim 100$\,km\,s$^{-1}$, although probably not as low as $\sim 10$\,km\,s$^{-1}$ (T.\ Janka 2007, private communication); these may be naturally provided during ECS events because of the short duration of the explosion limiting the development of asymmetries and build-up of larger kicks. 
\noindent
(3) NS recycling back to $\sim$\,ms spin periods occurs through long-term, steady accretion through Roche-lobe overflow. For some of the known double NS systems such mass transfer from the helium-rich progenitor of the second NS is actually required; however, depending on the dominant formation channel(s), mild recycling to $\sim10$\,ms may be possible (or even required) during the common envelope phase or through wind accretion from the helium-rich NS companion. 
\noindent
The diversity is apparent and it is providing us with new clues for the formation and origin of NS (see also \citep{S2008} for a related discussion).

In the near future we look forward to more analyses that can hopefully constrain some of the current uncertainties: what is the progenitor mass range for NS formation through both iron-core collapse and ECS, especially in binaries over a range of metallicities? Is there a minimum and/or maximum NS kick associated with ECS events? Do NS survive hypercritical accretion in common envelopes? Which progenitor properties determine the birth NS masses?

\begin{figure}
  \includegraphics[height=.29\textheight]{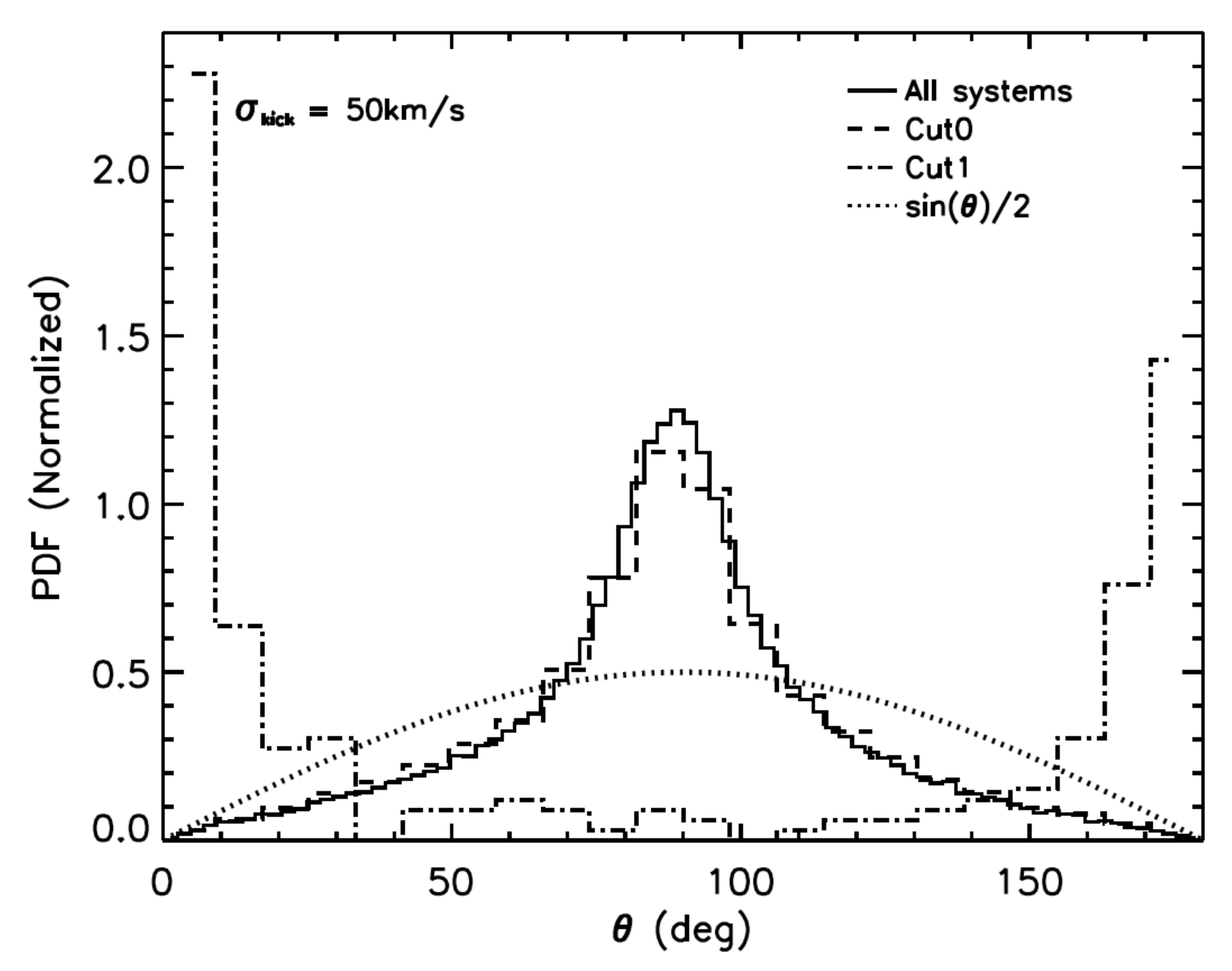}  
  \includegraphics[height=.285\textheight]{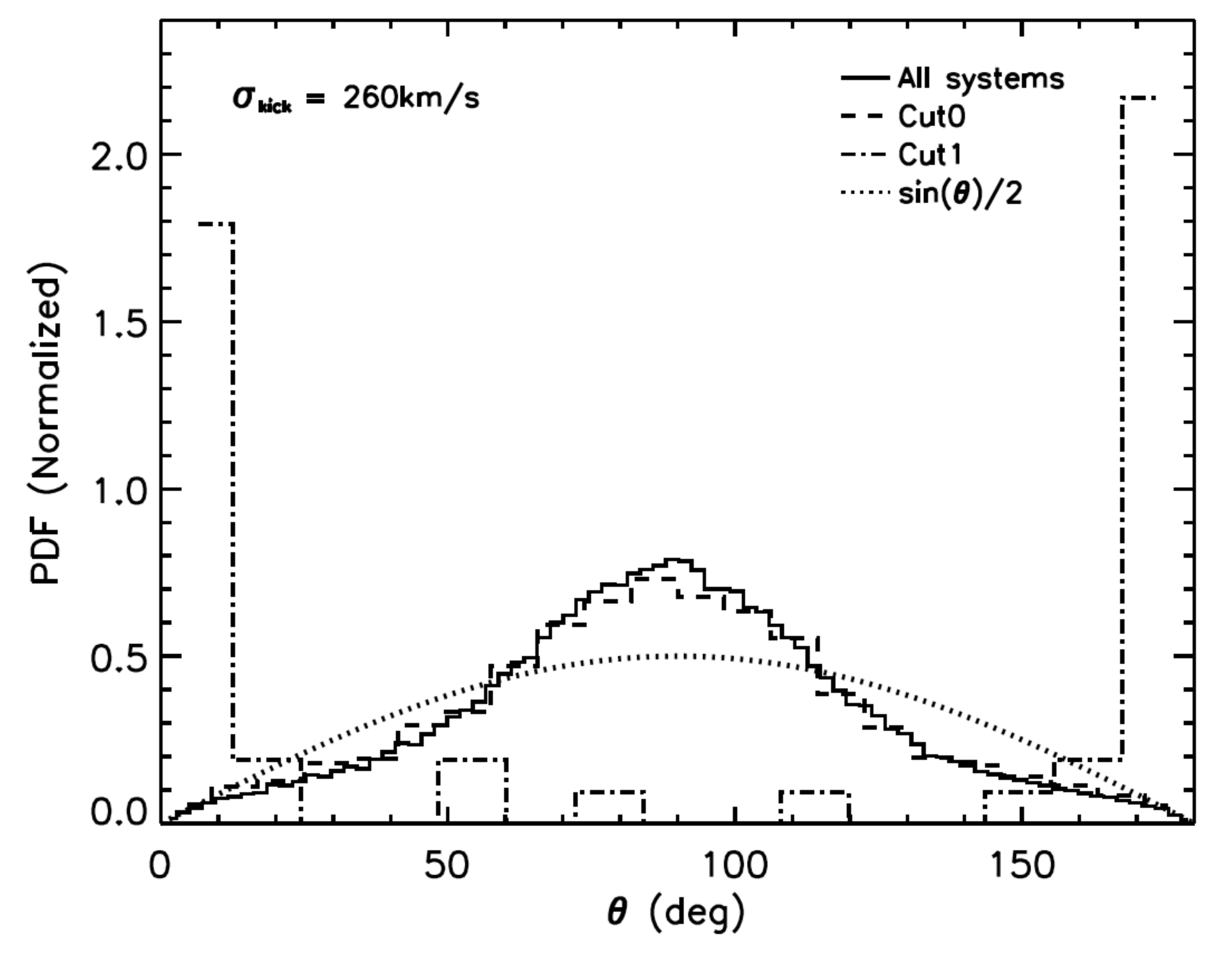} 
  \caption{Present-day distributions of the direction angle $\theta$ between NS-NS systemic velocities and our line-of-sight, for population synthesis models adopting Maxwellian kick velocity distributions with dispersions of 50\,km\,s$^{-1}$ (left) and 260\,km\,s$^{-1}$ (right). Solid lines represent the full population of NS-NS binaries, dashed lines the sub-population of systems within 50\,pc from the Galactic plane (Cut0), and dot-dashed lines the sub-population of systems within 50\,pc from the Galactic plane with a proper motion of less than 10\,km\,s$^{-1}$ (Cut1). For comparison, the isotropic $\theta$-distribution $f(\theta) = (\sin \theta)/2$ is also shown (dotted line). All PDFs are constructed using the NS-NS populations synthesized by Willems et al. (2006).}
  \label{f1}
\end{figure}
\begin{figure}
  \includegraphics[height=.29\textheight]{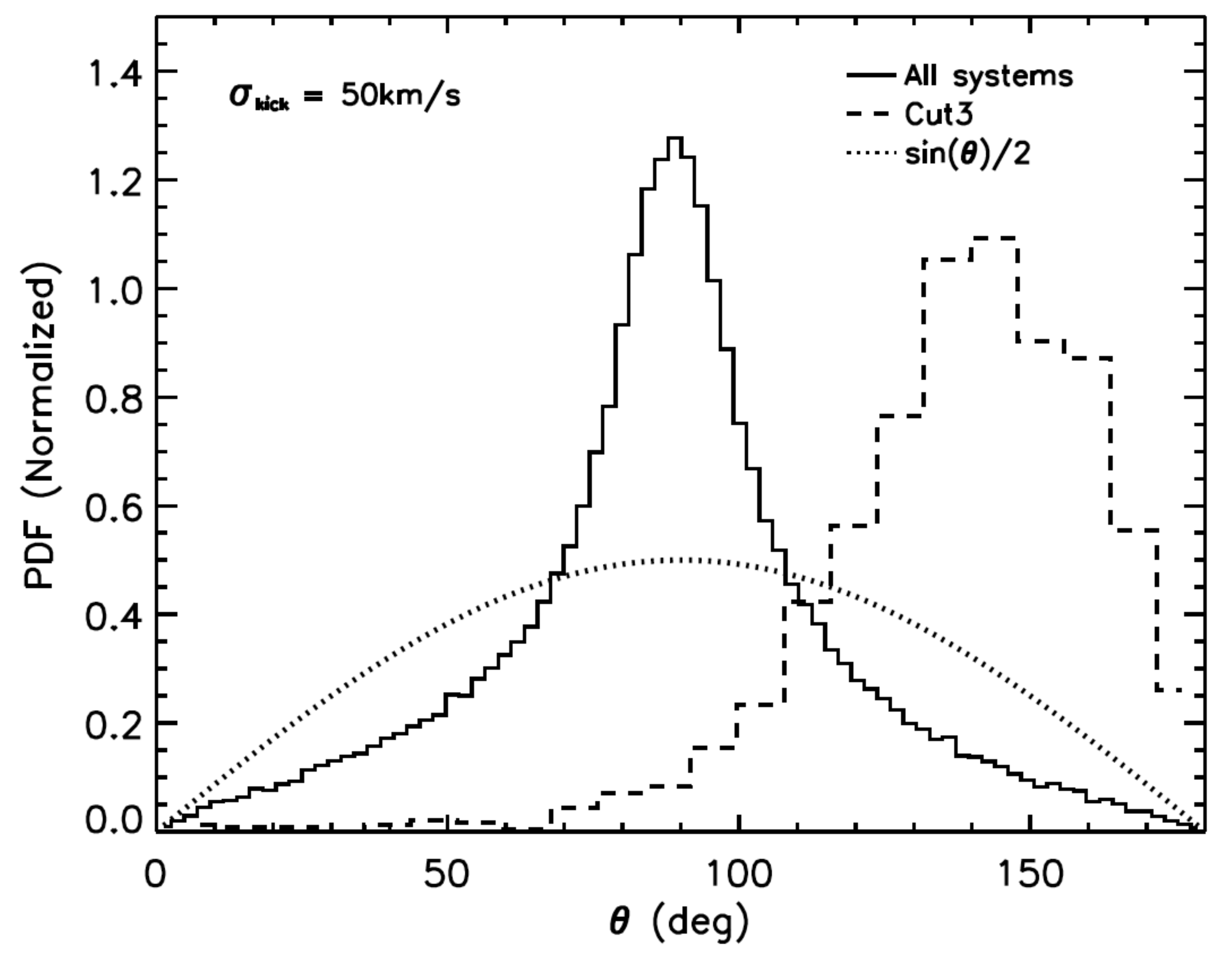} 
  \includegraphics[height=.29\textheight]{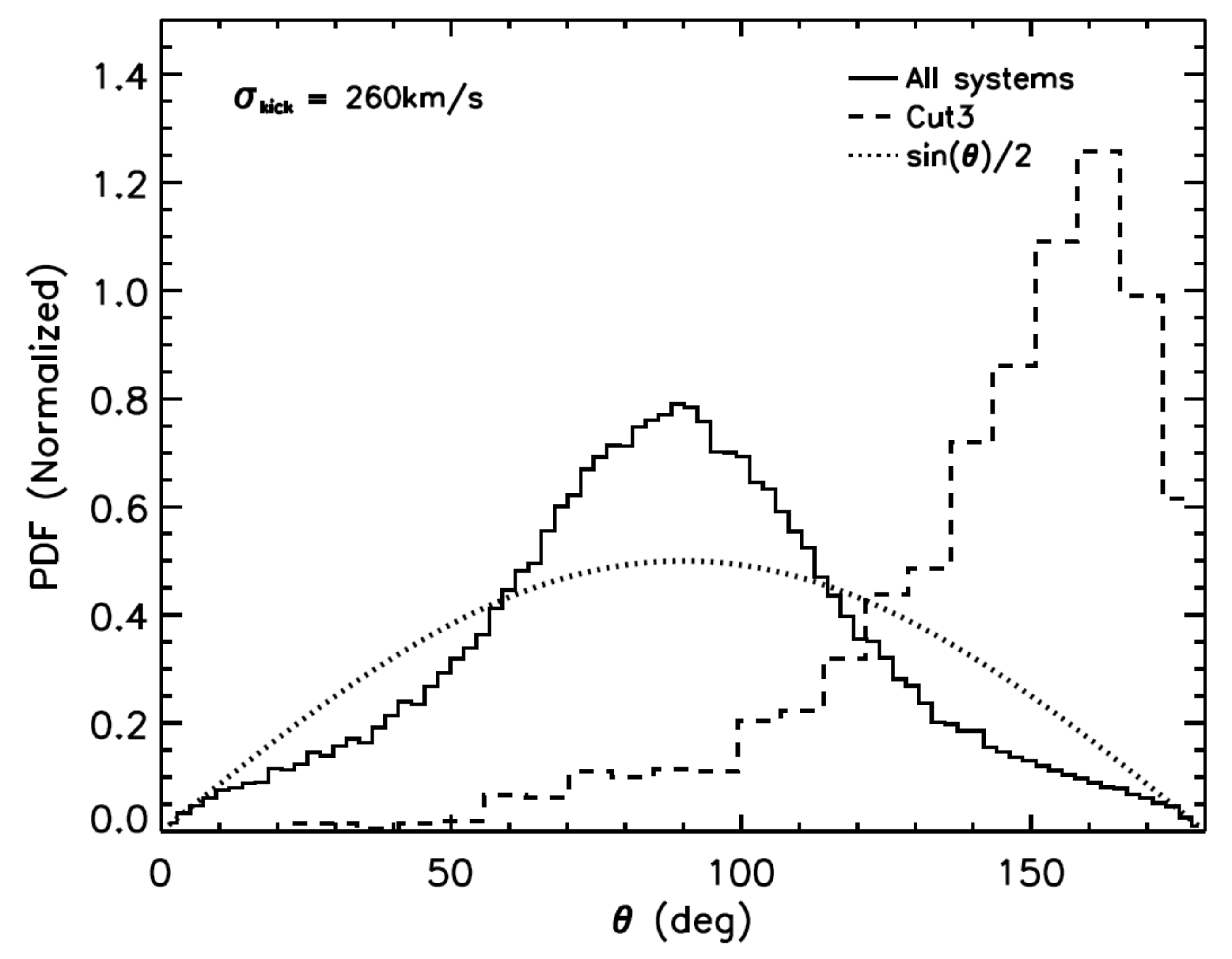} 
  \caption{As Figure~\ref{f1}, but with different constraints imposed on the population. Solid lines represent the full population of NS-NS binaries (as in Figure ~\ref{f1}), and dashed lines the sub-population of systems located within $\pm 20^\circ$ from PSR J0737-3039's position in Galactic longitude and latitude ($l=245.2^\circ$, $b=-4.5^\circ$; Cut3). The dotted line represents the isotropic $\theta$-distribution $f(\theta) = (\sin \theta)/2$. }
  \label{f2}
\end{figure}

%%%%%%%%%%%%%%%%%%%%%%%%%%%%%%%%%%%%%%%%%%%%%%%%
%% BACKMATTER
%%%%%%%%%%%%%%%%%%%%%%%%%%%%%%%%%%%%%%%%%%%%%%%%

\begin{theacknowledgments}
We are grateful to Ingrid Stairs for useful discussions and comments on this manuscript, and to Jeff Kaplan and Chris Belczynski for allowing us use of their codes from their earlier work in collaboration with VK and BW. This work is partially supported by a Packard Foundation Fellowship in Science and Engineering and a NSF CAREER grant (AST-0449558) to VK. 
\end{theacknowledgments}

\section{Appendix} 

All of the analyses of evolutionary histories and kinematics of pulsar systems suffer from the lack of any constraints on their radial velocities \citep{ 2004ApJ...616..414W, 2005ApJ...619.1036T, 2000ApJ...528..401W,2006PhRvD..74d3003W, 2006MNRAS.373L..50S}. In the case of PSR~J0737-3039, probably the most exhaustively studied binary pulsar, the question of the appropriate radial velocity distribution that should be adopted is one of the most important remaining uncertainties in the models (along with the age). Piran \& Shaviv \citep{2005PhRvL..94e1102P, 2005astro.ph.10584P} did not assume anything, but therefore had to restrict the Galactic motion to one dimension; Willems et al. \citep{2004ApJ...616..414W} assumed a broad flat radial velocity distribution; Willems et al. \citep{2006PhRvD..74d3003W} used population synthesis for the formation and motion of NS-NS binaries to derive theoretical radial velocity distributions filling in the void left by the lack of empirical constraints; Stairs et al. \citep{2006MNRAS.373L..50S} also adopted the latter in combination with updated proper motion constraints, but at the same time argued in favor of a radial velocity distribution obtained by assuming an isotropic distribution for the angle $\theta$ between the full 3-D space velocity and our line-of-sight. The choice between these two priors alters their most likely kick velocity at 95\% confidence level from 45--1005\,km\,s$^{-1}$ to 20--140\,km\,s$^{-1}$. Given this level of influence on results, we examine this question in more detail in what follows. 

The isotropic assumption appears reasonable in the absence of other constraints and most importantly appears intuitive. Quantitatively, it leads to a much narrower radial velocity PDF than the one obtained from population synthesis models (see Figure~1 in \citep{2006MNRAS.373L..50S}	). However, it is possible that the motion of the NS-NS population introduces velocity anisotropicities, especially when proper motion and location constraints are incorporated. In particular, if the motion is dominated by Galactic rotation, as it is for the progenitors of NS-NS, the systemic velocity distribution will be strongly anisotropic and depend on the direction of the line-of-sight. Specifically let us consider the following {\em thought experiment}: if only small kicks are imparted to the second-born NS at birth, an observer located at the Galactic center will see a $\theta$-distribution that is {\em strongly peaked} near angles of $90^\circ$. It is hard to envision that such a distribution would appear isotropic when the observer's viewpoint is shifted from the Galactic center to the position of the Sun.

To illustrate this further, we use the theoretical NS-NS populations synthesized by Willems et al. \citep{2006PhRvD..74d3003W} in the case of isotropic SN kick velocities with magnitudes drawn from Maxwellian distributions with dispersions of 50\,km\,s$^{-1}$ (as an example of pre-dominantly low kicks) and 260\,km\,s$^{-1}$ (as inferred by Hobbs et al.\ \citep{2005MNRAS.360..974H}).   We follow the formation and motion of these systems in the Galaxy and use the results to derive present-day radial and transverse velocities of NS-NS binaries as well as PDFs for the angle $\theta$ between the 3-D space velocities and the line-of-sight. We note that with this definition for $\theta$ the radial velocity $V_r$ is related to the transverse velocity $V_t$ as $V_r = V_t/\tan \theta$, so $\theta=0$ points away from us and $\theta=\pi$ towards us. Stairs et al. \citep{2006MNRAS.373L..50S} used this relation to derive an empirical radial velocity distribution for PSR J0737-3039 using $V_t = 10\,{\rm km\,s^{-1}}$ \citep{2006Sci...314...97K} and assuming an isotropic distribution for $\theta$. 

In Figure~\ref{f1}, we show the resulting present--day PDFs of the angle $\theta$ for the entire NS-NS population in the two kick velocity models, for the sub-population of systems within 50\,pc from the Galactic plane, and for the sub-population of systems within 50\,pc from the Galactic plane with a proper motion of less than10\,km\,s$^{-1}$. The distributions are {\em clearly anisotropic} for both kick velocity models, with a strong preference for directions along the line-of-sight (towards or away from us) emerging when the population of NS-NS systems is restricted to binaries within 50\,pc from the Galactic plane {\em and} a small proper motion of less than 10\,km\,s$^{-1}$, characteristic of PSR J0737-3039. In Figure~\ref{f2}, we show the $\theta$-PDFs for the sub-population of NS-NS binaries located within $\pm 20^\circ$ from PSR J0737-3039's position in Galactic longitude and latitude. Here too, the distribution of $\theta$-angles is clearly anisotropic, with velocity vectors pointing primarily towards us. The anisotropy persists if the sample population is restricted to systems within 1\,kpc from the Sun (the current upper limit on the distance from the Sun to PSR J0737-3039 \citep{2006Sci...314...97K}). The same holds true when combining the constraints on the vertical distance to the Galactic plane, the proper motion, and the position in terms of Galactic coordinates, though the statistics of the simulated samples become considerably poorer. 

We conclude that, although it appears intuitive that the direction of the space velocity of the double pulsar is distributed isotropically around the line-of-sight, the system's motion due to Galactic rotation and the post-SN peculiar velocity combined with the non-central position of observers on Earth induce significant anisotropies in the  double pulsar's most likely space velocity. Such anisotropies are implicitly taken into account in the theoretical radial velocity distributions derived and used by Willems et al. \citep{2006PhRvD..74d3003W} to constrain the formation of the second-born pulsar in PSR J0737-3039; therefore the resulting broader distributions of radial velocities are physically motivated and self-consistent compared to the narrower distribution obtained with the ad hoc assumption of isotropic velocities with respect to our line-of-sight.

%% The code below assumes that BibTeX is used. Compliant BibTex styles
%% are aipproc (for use with natbib) and aipprocl (if natbib is missing
%% at the site).
%%
%% Please run "bibtex \jobname" to obtain the bibliography and 
%% then re-run LaTeX twice to fix the references!
%%
%% When referring to citations in the text, in quare brackets [] show
%% the number in order of appearance. References in the References
%% section are listed in the same numerical order.
%%%%%%%%%%%%%%%%%%%%%%%%%%%%%%%%%%%%%%%%%%%%%%%%%

\bibliographystyle{aipproc}   % if natbib is available

\end{document}